\documentclass[sigconf]{acmart}

\usepackage[caption=false]{subfig}
\usepackage{booktabs}
\usepackage[utf8]{inputenc}
\usepackage{multirow}
\usepackage{array}
\usepackage{listings}

\setcopyright{acmcopyright} 
\copyrightyear{2017} 

\renewcommand\footnotetextcopyrightpermission[1]{}

\newcolumntype{C}{>{\centering\arraybackslash}m{2cm}}
\newcolumntype{E}{>{\centering\arraybackslash}m{4.5cm}}
\newcolumntype{F}{>{\centering\arraybackslash}m{1cm}}
\newcolumntype{G}{>{\centering\arraybackslash}m{1.5cm}}
\newcolumntype{H}{>{\centering\arraybackslash}m{0.85cm}}

\begin{document}

\title[Classifying the Correctness of Generated White-Box Tests]{Classifying the Correctness of Generated White-Box Tests: An Exploratory Study}

\author{David Honfi}
\orcid{0000-0001-5217-828X}
\affiliation{%
  \institution{Budapest University of Technology and Economics}
  \city{Budapest} 
  \state{Hungary}
}
\email{honfi@mit.bme.hu}

\author{Zoltan Micskei}
\orcid{0000-0003-1846-261X}
\affiliation{%
  \institution{Budapest University of Technology and Economics}
  \city{Budapest} 
  \state{Hungary}
}
\email{micskeiz@mit.bme.hu}

\begin{abstract}
White-box test generator tools rely only on the code under test to select test inputs, and capture the implementation's output as assertions. If there is a fault in the implementation, it could get encoded in the generated tests. Tool evaluations usually measure fault-detection capability using the number of such fault-encoding tests. However, these faults are only detected, if the developer can recognize that the encoded behavior is faulty. We designed an exploratory study to investigate how developers perform in classifying generated white-box test as faulty or correct. We carried out the study in a laboratory setting with 54 graduate students. The tests were generated for two open-source projects with the help of the IntelliTest tool. The performance of the participants were analyzed using binary classification metrics and by coding their observed activities. The results showed that participants incorrectly classified a large number of both fault-encoding and correct tests (with median misclassification rate 33\% and 25\% respectively). Thus the real fault-detection capability of test generators could be much lower than typically reported, and we suggest to take this human factor into account when evaluating generated white-box tests.
\end{abstract}



\maketitle


\section{Introduction}
\label{sec:introduction}

Due to the ever increasing importance of software, assessment of its quality is essential. In practice, \emph{software testing} is one of the most frequently used techniques to improve software quality. Thorough testing of software demands significant time and effort. To alleviate the tasks of testers and developers, several \emph{automated techniques} have been proposed \cite{anand-testgen-survey}. These advanced methods are often available as off-the-shelf tools, e.g., Pex/IntelliTest \cite{pexwb}, Randoop \cite{pacheco-randoop}, or EvoSuite \cite{fraser-evosuite}. Some of these techniques can rely \emph{only on the source/binary code} to select relevant inputs. For the selected inputs these white-box test generators record the implementation's actual output in test asserts. However, if only the implementation is used, the \emph{assertions} created in the generated test cases contain the \emph{observed behavior}, not the \emph{expected}. 

As these techniques and tools evolve, more and more empirical evaluations are required to assess their usefulness. In most of the studies, the tools were evaluated in a technology-oriented setting (e.g., \cite{kracht-experiment,wang-klee-manual,shamshiri-experiment}). Only a limited number of studies involved \emph{human participants} performing prescribed tasks with the tools \cite{fraser-experiment-extended,rojas-experiment,enoiu-experiment}.

A common aspect to evaluate the effectiveness of test generator tools  is the \emph{fault detection capability} of the generated tests. Related studies~\cite{manual-generated-industrial,fraser-experiment-extended,bugfixing-empirical,rojas-experiment,ramler-experiment-legacy,shamshiri-experiment,Nguyen:2013} typically employ two metrics for this purpose: 1) mutation score or 2) number of detected faults. Although mutation score has been shown to be in correlation with real fault detection capability \cite{mutants-real-fault}, it has concerns to be aware of \cite{Papadakis-2016}. The number of detected faults is usually measured using a faulty version (with injected faults) and a fault-free version (original) of the code under test. If a generated test passes on the faulty version and fails on the original, it is considered as a \emph{fault-detector}.

\begin{figure}[ht]
    \centering
    \includegraphics[width=0.9\columnwidth]{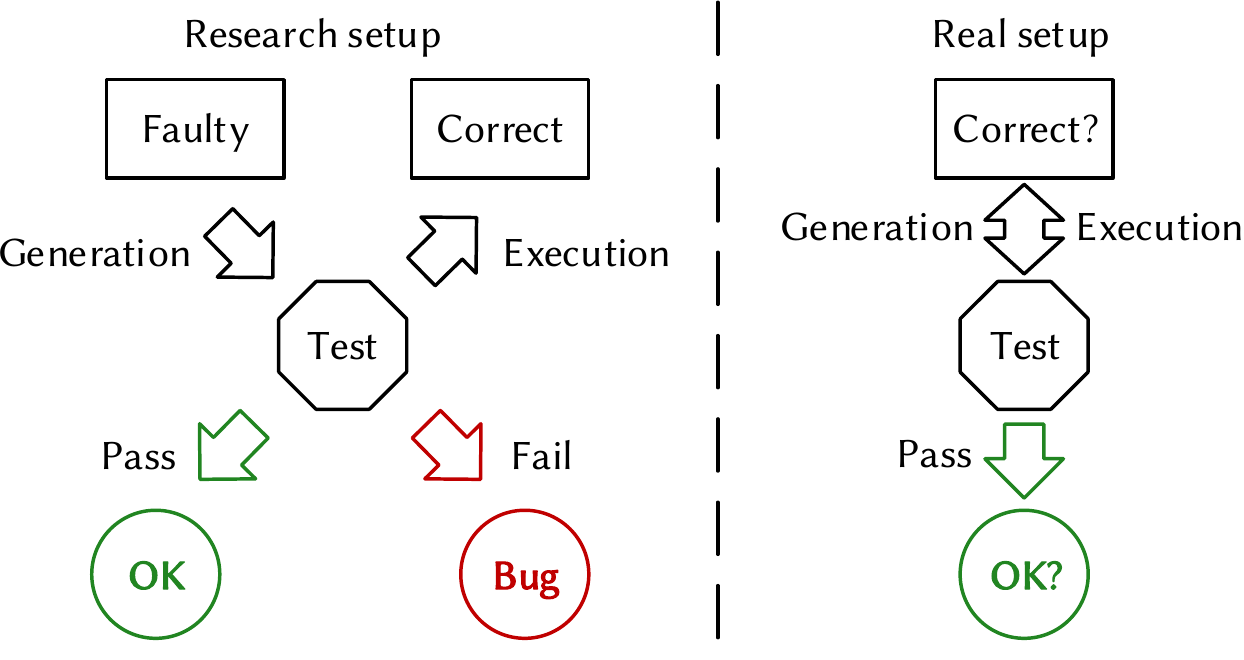}
    \caption{The problem of evaluating generated tests.}
    \label{fig:problem-overview}
\end{figure} 

However, a fundamental problem is that we have \emph{no a priori knowledge} about the correctness of the implementation in real scenarios (i.e., there is no faulty and correct version, see Figure~\ref{fig:problem-overview}). If the test generator uses only the program code, then the user of the test generator \emph{must validate each assertion} in the generated test code to decide whether the test encodes an expected or a faulty behavior. Although this is very simple for trivial errors, it could be rather complex in case of slight mismatches between the implementation and its intended behavior. Note that the number of generated tests to examine could be decreased with implicit or derived oracles~\cite{oracle-problem} (e.g., robustness or regression testing), but if the generated tests are used for functional testing, then in the end some of the validation needs to be performed by the developers and testers. However, it is not evident that humans can correctly identify all faults that can be possibly detected using the generated tests. Although some experiments (e.g., Fraser et al.~\cite{fraser-experiment-extended}) mentioned this potential issue, most of the related studies do not consider it as a validity threat during their evaluations. The \emph{consequence} of this is that the practical fault-finding capability of the test generators \emph{can be much lower} than presented in experimental evaluations.

Thus the question that motivated, and served as a basis of our research is the following: \emph{How do developers perform in using the tests generated from code to detect faults and decide whether the implementation is correct? \footnote{Note that if a test generated from a faulty implementation encodes a fault but passes, then the test can be considered faulty as well. Therefore classifying the tests as faulty or correct could reveal a faulty implementation.}} This question is mainly motivated by the fact that the actual fault-finding capability of white-box test generator tools could be much lower than reported in already existing experiments due to the classification performance of tool users.

We designed and performed an \emph{exploratory study} with human participants that covers a realistic scenario resembling developers testing previously untested code with the help of test generators. The participants' task was to classify tests generated by Microsoft IntelliTest \cite{pexwb} whether they encode faulty or correct behavior for two open source projects carefully selected from GitHub. The activities of participants were recorded using both logging and screen capture, and were analyzed quantitatively and qualitatively by coding the observed behaviors in each of the videos. 

Our results show that deciding whether a test encodes faulty behavior is a challenging task even in a laboratory setting. Only 2 of the 54 participants were able to classify all 15 tests correctly, and the median of misclassification rate reached 33\% for fault-encoding tests. Surprisingly, a large number of correct tests were also classified as faulty (misclassification rate 25\%). The time required to classify one test case varied largely, but on average classifying one test required 2 minutes. Finally, the effect of the experience of the participants on the classification performance was analyzed.

In experimental research, replications and secondary studies are vital to increase the validity of results. Thus we made the dataset, the videos, the coding of behavior and the full analysis scripts available for further use \cite{dataset}. 

The main contributions of the paper are as follows.

\begin{itemize}	
	\item We \emph{designed} a new exploratory study with human participants to investigate the importance of the classification of generated tests (Sect.~\ref{sec:experiment-design}).
	\item We \emph{performed} the study with 54 participants (Sect.~\ref{sec:execution}) and \emph{analyzed} the results (Sect.~\ref{sec:results}) showing \emph{evidence} that classification of generated tests is not easy and humans can not necessarily detect all faults.
	\item We \emph{drew conclusions} from the results and \emph{gave recommendations} for further studies and replications (Sect.~\ref{sec:discussion}).
\end{itemize}


\section{Related work}
\label{sec:motivation}

\paragraph{Test generation and oracles} Anand et al.~\cite{anand-testgen-survey} present a survey about test generation methods, including those that generate tests only from binary or source code. As these methods do not have access to a  specification or model, they rely on other techniques than specified test oracles~\cite{oracle-problem}. For example, for certain outputs \emph{implicit oracles} can be used: a segmentation fault is always a sign of a robustness fault~\cite{Shahrokni20131}, while finding a buffer overflow means a security fault~\cite{sage-2013}. Other implicit oracles include general contracts like \verb|o.equals(o)| is true~\cite{pacheco-randoop}. However, test generators usually generate numerous tests passing these implicit oracles. For handling these tests there are basicly two options. On the one hand the developer could specify \emph{domain-specific partial specifications} (e.g., as parameterized tests~\cite{pexwb} or property-based tests~\cite{quickcheck}). On the other hand the tools usually record the \emph{observed output} of the program for a given test input in assertions, and the developer could \emph{manually examine} these asserts to check whether the observed behavior conforms to the expected behavior.

In our paper we consider this latter case, i.e., where there is no automatically processable full or partial specification, the generated tests were already filtered by the implicit oracles, and the remaining tests all passed on the implementation, but we cannot be sure if they encode the correct behavior. In this case \emph{derived oracles} are commonly used to \emph{decrease} the number of tests to manually examine or ease the validation. For example, existing tests can be used to generate more meaningful tests~\cite{human-oracle-cost},  similarity between executions can be used to pinpoint suspicious asserts~\cite{Pastore:2015}, or clustering techniques can be used to group potentially faulty tests~\cite{Almaghairbe2016}. Moreover, if there are \emph{multiple versions} from the implementation (e.g., regression testing~\cite{yoo-survey} or different implementations for the same specification~\cite{pacheco-randoop}), tests generated from one version could be executed on the other one. However, even in this scenario, tests do not detect faults, but merely \emph{differences} that need to be manually inspected (e.g., a  previous test can fail on the new version not because of a fault, but because a new feature has been introduced). In summary, none of these techniques can classify all tests perfectly, and the remaining ones still need to be examined by a human. 

\paragraph{Testing studies involving participants} Juristo et al.~\cite{juristo-25-years-testing} collected testing experiments in 2004, but only a small number of the reported studies involved human subjects (e.g., Myers et al.~\cite{myers-experiment}, Basili et al.~\cite{basili-experiment}). More recently, experiments evaluating test generator tools were performed: Fraser et al.~\cite{fraser-experiment-extended} designed an experiment for testing an existing unit either manually or with the help of EvoSuite; Rojas et al.~\cite{rojas-experiment} investigated using test generators during development; Ramler et al.~\cite{ramler-experiment-legacy} compared tests written by the participants with tests generated by the researchers using Randoop; and Enoiu et al.~\cite{enoiu-experiment} analyzed tests created manually or generated with a tool for PLCs. These experiments used mutation score or correct and faulty versions to compute fault detection capability.

\paragraph{Related studies} We only found two studies that are closely related to our objective. In the study of Staats et al.~\cite{staats-understanding} participants had to classify invariants generated by Daikon. They found that users struggle to determine the correctness of generated program invariants (that can serve as test oracles). The object of the study was one Java class, and tasks were performed on printouts. Pastore et al.~\cite{crowdoracles} used a crowd sourcing platform to recruit participants to validate JUnit test cases based on the code documentation. They found that the crowd can identify faults in the test assertions, but misclassified several harder cases. These studies suggest that classification is not trivial. Our study extends these results by investigating the problem in a setting where participants work in a development environment on a more complex project.


\section{Study planning}
\label{sec:experiment-design}

\begin{table*}[ht]
    \centering
    \caption{Details of the selected objects (projects and classes under test).}
    \label{table:selected-objects}    
    \begin{tabular*}{\textwidth}{ lccllcccp{6cm} }\toprule 
        \multicolumn{3}{c}{\emph{Project}} & & \multicolumn{4}{c}{\emph{Selected class}} \\
        \cmidrule{1-3} \cmidrule{5-9}
        \emph{Name} & \emph{\#Classes} & \emph{KLOC} & & \emph{Name} & \emph{\#Methods} & \emph{\#Dependencies} & \emph{LOC} & \emph{Selected methods} \\
        \midrule
        NBitcoin & 602 & 29.8 & & AssetMoney & 44 & 2 & 202 & CompareTo, Constructor, Equals, Min, Plus \\
        Math.NET & 271 & 48.7 & & Combinatorics & 17 & 4 & 95 & Combinations, CWithRepetition, Permutations, Variations, VWithRepetition \\
        \bottomrule
    \end{tabular*}
\end{table*}

\subsection{Goal and method}

Our main goal was to study whether developers can use and validate the tests generated only from program code by classifying whether a test encodes a correct behavior or a fault. 

As there is little empirical evidence about the topic, to understand it better we followed an exploratory and interpretivist approach~\cite{Wohlin}. We formulated the following base-rate and relationship \emph{research questions}~\cite{selecting-empirical-methods}.

\begin{description}
	\item[RQ1] How do users of white-box test generation perform in the classification of generated tests?
	\item[RQ2] How much time do users spend with the classification of generated tests?
	\item[RQ3] What could impact the users' ability to correctly classify generated test cases?
\end{description}

Note that RQ3 is intentionally defined so that it requires a mix of exploratory and post-hoc analyses.

As these test generator tools are not yet widespread in industry we selected an \emph{off-line context}. We employed an exploratory study in a laboratory setting using students as human participants. Our research process involved both qualitative and quantitative phases. We collected data using both observational and experimental methods. The data obtained was analyzed using exploratory data analysis and statistical methods, and by coding behaviors in screen capture videos. For the design and reporting of our study, we followed the guidelines of empirical software engineering \cite{selecting-empirical-methods,Wohlin,Wohlin-book}.

\subsection{Variable selection}

Understanding and validating generated tests is a rather complex task that can be affected by numerous variables. We focus on the following \emph{independent variables}. For each variable possible levels are listed, from which the bold ones are selected for our study.

\begin{itemize}
	\item \emph{Participant source:} Where the participants are recruited from [\textbf{students}, professionals, mixed].
	\item \emph{Participant experience:} Experience in testing and test generation tools [none, \textbf{basic}, experienced].
	\item \emph{Participant knowledge of objects:} Whether the participant has a priori knowledge about the implementation under test [known, \textbf{unknown}].
	\item \emph{Objects source:} The source, where the objects are selected from [\textbf{open source}, closed source, artificial/toy, \ldots].
	\item \emph{Object source code access:} Whether the objects are fully visible to the participants [\textbf{white-box}, black-box].
    \item \emph{Object versions:} How many versions are available for the code to test [\textbf{one}, multiple versions].
	\item \emph{Faults type:} The source and type of the faults used in the objects [real, \textbf{artificial}, mutation-based].
	\item \emph{Specification type:} How the specifications of the objects are given [\textbf{code comments}, text document, formal, \ldots].
	\item \emph{Test generator tool:} Which test generator is used for generating tests [\textbf{IntelliTest}, EvoSuite, Randoop, \ldots].
	\item \emph{User activity:} The allowed user activities in the study [\textbf{run}, \textbf{debug}, modify code \ldots].
\end{itemize}

The following \emph{dependent variables} are observed: 

\begin{itemize}
    \item Answers of participants: Classification of each test as OK (correct) or wrong (faulty).
    \item Activities of participants: What activities do the participants perform during the task (e.g., running tests).
    \item Time spent by participants: How much time do the participants spend on each individual activity.
\end{itemize}

Note that as this is exploratory research there is no hypothesis yet, and because the research questions are not causality or comparative questions, all independent variables had fixed levels (i.e., there are no factors and treatment).

\subsection{Subjects (Participants)}

Our goal was to recruit people, who are already familiar with the concepts of unit testing and white-box test generation. We recruited participants from MSc students enrolled in one of our V\&V university course. They were suitable candidates as the course has covered testing concepts, test design, unit testing and test generation  ($5\times2$ hours of lectures, $3\time2$ hours of laboratory exercises and approximately 20 hours of group project work).

Participation in the study was optional, but we motivated it with giving extra points (approximately 5\% in the final evaluation of the course) for participation. However, we emphasized that these points were given independently from the experiment results not to have any negative performance pressure.

\subsection{Test generator tool}

As our V\&V course has laboratory exercises with the IntelliTest test generation tool, we choose this tool for the study as well. IntelliTest (formerly known as Pex~\cite{pexwb}) is a state-of-the-art dynamic symbolic execution-based test generator. IntelliTest currently supports the C\# language and is integrated into Visual Studio 2015. IntelliTest's basic concept is the \emph{parameterized unit test}, which is a test method with arbitrary parameters called from the generated test cases with concrete arguments.

\subsection{Objects (Projects and classes)}

The main requirements towards the objects were that they should be written in C\#, IntelliTest should be able to explore them, they should be not too complex so that participants could understand them during the task, but they should contain multiple non-trivial classes depending on each other, otherwise faults could be easily identified just by reviewing the code of the class under test. We did not find projects satisfying these requirements in previous studies, thus we searched for open source projects.

Our project selection criteria included the followings.

\begin{itemize}
	\item Shall have at least 500 stars on GitHub: this likely indicates a project that really works and filters out prototypes and not working code.
	\item Should not have any relation to graphics, user interface, multi-threading, multi-platform execution: these may introduce difficulties for the test generator tool.
	\item Shall be written in C\# language: IntelliTest only supports this language.
	\item Shall be compilable in a few seconds: this makes users able to run fast debugging sessions during the experiment.
\end{itemize}

We decided to use \emph{two different classes from two projects} with vastly different characteristics. The selection criteria for the classes were the followings.

\begin{itemize}
	\item Shall be explorable by IntelliTest without issues to have usable generated tests.
	\item Shall have more than 4 public methods to have reasonable amount of generated test cases.
	\item Shall have at least 1 external invocation pointing outside the class, but not more than 3. This ensures a fault-injection location for the experiment.
	\item Shall have at least partial commented documentation to use as specification.
\end{itemize}

Based on pilots, we found that participants can examine \emph{15 tests} in a reasonable amount of time. To eliminate the bias possibly caused by tests for the same methods, we decided to have the 15 cases \emph{for 5 different methods}.

\subsubsection{Project and class selection}

Finding suitable objects turned out to be much harder than we anticipated. We selected 30 popular projects from GitHub as candidates that seemed to satisfy our initial requirements. However, we had to drop most of them: either they heavily used features not supported by IntelliTest (e.g., multi-threading or graphics), or did not have inter-class dependencies, or would have required extensive configuration (e.g., manual factories, complex assumptions) to generate non-trivial test values.

Finally we kept the two most suitable projects: 

\begin{itemize}
    \item \emph{Math.NET Numerics} \cite{mathnet} is a .NET library that offers numerical calculations in probability theory or linear algebra. It contains mostly data structures and algorithms.
    \item \emph{NBitcoin} \cite{nbitcoin} is a more business-like library, which is available as the most complete BitCoin library for .NET.
\end{itemize}

Table~\ref{table:selected-objects} lists the selected classes and methods of the two projects. The \verb|Combinatorics| class implements enumerative combinatorics and counting: combinations, variations and permutations, all with and without repetitions. The \verb|AssetMoney| class implements the logic of the Open Asset protocol for arbitrary currencies that have conversion ratio to BitCoin.

Most of the selected methods had method-level comments originally containing the description of correct behavior, but we extended them slightly based on the feedback from pilots. They are still not perfect, but they represent comments used in real projects.

\subsubsection{Fault selection and injection}

To obtain fault-encoding tests from IntelliTest, faults need to be injected into the objects. There are several alternatives to obtain such faults, each of them would affect the validity of the study. As we were not able to extract meaningful faults for the selected classes from the version history of the project, we used \emph{artificial faults} in a systematic way. We selected different, representative fault types \cite{faults-survey} from the Orthogonal Defect Classification~\cite{odc-classification}. The cited survey identifies the most commonly committed fault types in real-world programs. We used this survey as the source of the selected the faults: we selected each from the top quarters of the ODC categories (see Table~\ref{table:selected-faults}). During the injection procedure we made sure that the faults 1) have no cross-effects on each other, and 2) have no effect on behavior other than the intented. We injected three faults in both projects. In case of NBitcoin we injected the faults inside the selected class, while in case of Math.NET into the project's other classes.

\begin{table*}[!ht]
	\centering
    \setlength{\tabcolsep}{3pt}
   	\caption{Selected faults and their realizations in the code.}
    \label{table:selected-faults}    
	\begin{tabular*}{\textwidth}{ p{1.5cm} | F m{3cm} C E E }\toprule 
		Project & ID & Name & ODC category & Original snippet & Faulty snippet \\
		\midrule
		& F10 & Extraneous assigment using another variable & Assignment & - & \verb|dec = dec + divisibility;| \\
		NBitcoin & F3 & Missing function call & Function & \verb|return _Quantity| \verb|.Equals(other.Quantity);| & \verb|return true;| \\
		 & F6 & Algorithm - large modification & Function & \verb|return left.Quantity| \verb|<= right.Quantity;| & \verb|return left.Id._Bytes.Length| \verb|<= right.Id._Bytes.Length;| \\
		\midrule
		& F2 & Missing OR sub-expr in expression used as branch condition & Check & \verb!k < 0 || n < 0 || k > n! & \verb!k < 0 || n < 0! \\
		Math.NET & F5 & Wrong logical expression used as branch condition & Check & \verb!k < 0 || n < 0 || k > n! & \verb!k < 0 && n < 0 || k > n! \\
		& F8 & Wrong arithmetic expression in parameter of function call & Interface & \verb|SpecialFunctions| \verb|.FactorialLn(n + k - 1)| & \verb|SpecialFunctions| \verb|.FactorialLn(n - k + 1)| \\
		\bottomrule
	\end{tabular*} 
\end{table*}

\subsubsection{Generated tests}

We generated tests with IntelliTest for each selected method using parameterized unit tests. Tests were generated from the version already containing the above faults. There were methods, where IntelliTest could not generate values that cover interesting behaviors. In these cases, we extended the parameterized unit tests with special assumptions that request at least one test case from IntelliTest with values that fulfill the preconditions. From each test case set, we selected 3 test cases for the study. We choose the most distinct cases that cover vastly different behaviors in the method under test. Each test case was given an identifier ranging from 0 to 14 (therefore both NBitcoin and Math.NET have tests T0 to T14). Furthermore, the corresponding method is indicated with a suffix in each test case identifier. Thus for the first method, three cases were generated: T0.1, T1.1 and T2.1. IntelliTest generates one test file for each method, but we moved the test cases into individual files to help tracking the participants.

\subsection{Environment}

A Windows 7 virtual machine was used that contained the artifacts along with Visual Studio 2015 and Google Chrome. Participants were asked to use only two windows: 1) an experiment portal in Chrome, and 2) Visual Studio.

\begin{figure}[ht]
	\centering
	\includegraphics[width=0.7\columnwidth]{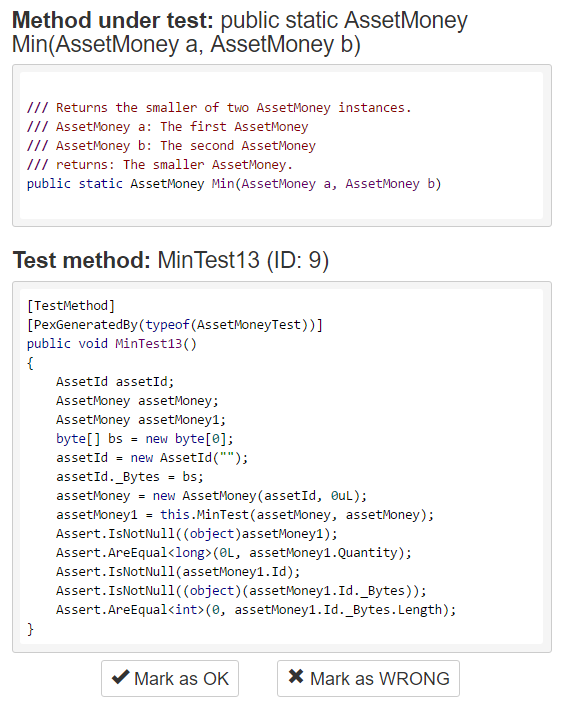}
	\caption{A test page in the experiment portal.}
	\label{fig:experiment-portal}
\end{figure}

We designed a special website, the experiment portal (Figure~\ref{fig:experiment-portal}) in order to record the answers of the participants. It was a more reliable way to collect the results than using some mechanism in the IDE (e.g., using special comments), as participants could un-intendedly delete or regenerate the test code.

Participants used this portal to decide whether the test case is wrong or correct with respect to the specification. The portal displayed the test code and the method comment of the corresponding method in the class under test. Participants could record their answer using two buttons. Participants could correct their already answered cases. Questions could be skipped if a participant was not sure in the answer (however, nobody used that option).

In Visual Studio the default development environment was provided with a simple activity tracking extension. Participants got the \emph{full project} with every class. Participants were asked 1) not to modify any code, 2) not to execute IntelliTest, and 3) not to use screen splitting. On the other hand, we encouraged them to \emph{use test execution} and \emph{debugging} to explore the code under test.

\subsection{Procedure}

The main procedure of the 2-hour session is as follows.

\begin{enumerate}
    \item Sign informed consent.
    \item Find a seat, receive a unique, anonymous identifier.
    \item Fill background questionnaire.
    \item Listen a to 10-minute overview presentation and go through a 15-minute guided tutorial.
    \item Perform the test \emph{randomly assigned} classification task in at most 1 hour.
    \item Fill an exit survey.
\end{enumerate}

Participants only receive one sheet of paper that describe both the procedure and the task with the path to the project and class under test. To obtain detailed knowledge about the participants, we designed a background questionnaire asking about the participants' experience with development and testing. Also, the questionnaire has a quiz in the end about C\# and testing.  In order to summarize the most important and required information, we designed a 10-minute presentation in which the procedure, the projects, the environment, the basic concepts of IntelliTest, and the rules are introduced. Furthermore, to make participants familiar with the environment and the task, a 15-minute guided tutorial is held on a simple project. This tutorial was specially elaborated to have both wrong and good answers for the test cases. During this tutorial, participants can ask anything. The main task is to classify each of the 15 generated test cases in the portal whether it is fault-encoding (wrong) or not (okay). Finally, an exit survey is filled that asks participants about their feelings regarding the task accomplished.

We planned to perform 2 study sessions as the room, where the study was planned to be conducted has only 40 seats available.

\subsection{Data collection}

We use two data collection procedures. On one hand, we extended the development environment so that it logs every window change, test execution and test debug as well. Also, we wrote a script that documents every request made to the experiment portal. On the other hand, we set up a screen recording tool to make sure that every action of the participants is recorded.

Each participant has 6 output files that is saved for data analysis.

\begin{itemize}
	\item \emph{Answers:} The answer submitted to the portal in JSON.
	\item \emph{Background:} The answers given in the background questionnaire in CSV format.
	\item \emph{Exit:} The answers given in the exit survey in CSV format.
	\item \emph{Portal log:} The user activity recorded in the portal.
	\item \emph{Visual Studio log:} The user activity recorded into a CSV-like format using a custom Visual Studio extension.
	\item \emph{Screen recorded video:} The participant activity during the main session in MP4 format.
\end{itemize}

\subsection{Data analysis}

First, the raw data is processed by checking the answers of the participants and coding the screen capture videos. Next, the processed data is analyzed using exploratory techniques and statistical tests.

\paragraph{Analysis of answers} We analyze the answers obtained from the experiment portal using binary classification for which the confusion matrix is found in Table~\ref{table:confusion-matrix}.

\begin{table}
	\centering
    \setlength{\tabcolsep}{3pt}
   	\caption{Confusion matrix for participant answers.}
    \label{table:confusion-matrix}
	\begin{tabular}{C p{2.5cm} p{2.5cm}}
	\toprule
    \emph{Correct answer}  & \multicolumn{2}{c}{\emph{Participant answer}} \\
    \cmidrule{2-3}
	                       & \emph{Marked as OK} & \emph{Marked as WRONG} \\
	\midrule
	OK              & true negative (TN)  &	false positive (FP)   \\
	WRONG           & false negative	(FN) &	true positive (TP)    \\
	\bottomrule
	\end{tabular}
\end{table}

\begin{table}
	\centering
    \setlength{\tabcolsep}{3pt}
   	\caption{Coding scheme of the video analysis.}
    \label{table:video-analysis}
    \begin{minipage}{\columnwidth}     
	\begin{tabular*}{\columnwidth}{ m{4.5cm} m{3cm}}\toprule 
	\emph{Behavior} & \emph{Modifiers} \\
	\midrule
	Portal activated & - \\
	Visual Studio activated & - \\
	Changed page in portal & T0-14, Home \\
	Change window in VS & CUT, SUT, PUT$^*$, T0-14 \\
	Marked as OK & - \\
	Marked as WRONG & - \\
	Remove answer & T0-14 \\
	Running test & - \\
	Debugging test & Start, End \\
	Submit & - \\
	\bottomrule
	\end{tabular*}
    \bigskip 
    
    \centering\footnotesize $^*$CUT: class under test, SUT: other system under test, PUT: parameterized unit test. 
    \end{minipage}
\end{table}

\begin{figure}[ht]
    \centering
    \includegraphics[width=1\columnwidth]{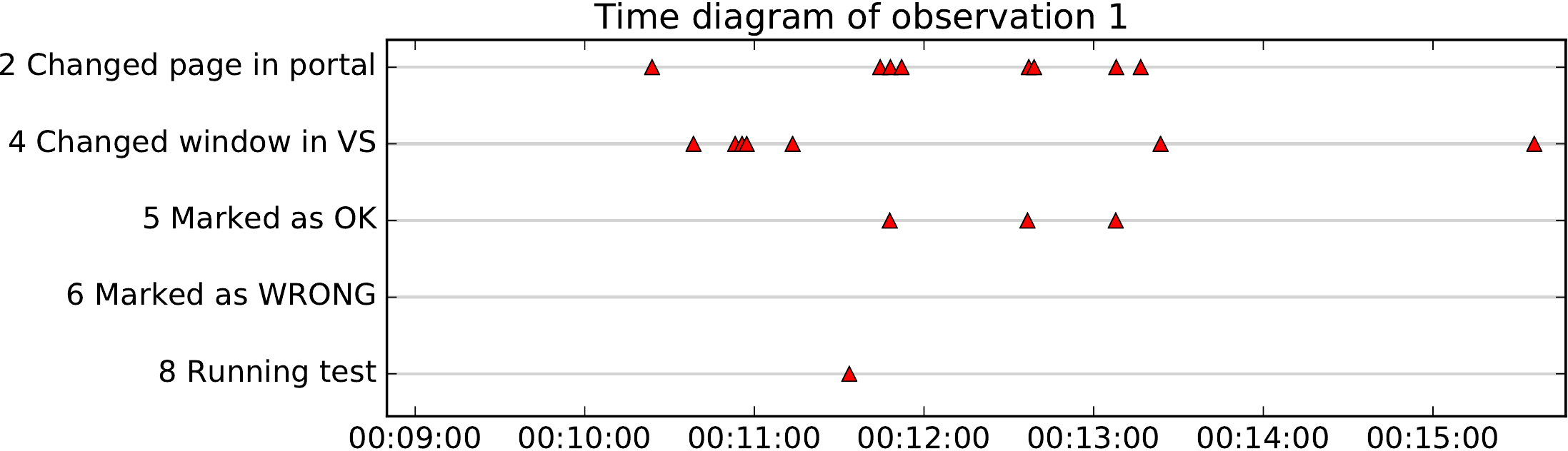}
    \caption{Coded events in Boris for one participant (excerpt).}
    \label{fig:boris-plot}
\end{figure}

\paragraph{Video coding}

We annotate every recorded video using an academic behavioral observation and annotation tool called Boris \cite{boris}. We designed a behavioral coding scheme that encodes every activity, which we are interested in. The coding scheme can be found in Table~\ref{table:video-analysis}, all occurrences of these events are marked in the videos. Note that, during the video coding, we only use point events with additional modifiers (e.g., change of page in the portal is a point event along with a modifier indicating the identifier of the new page). In order to enable interval events, we created modifiers with start and end types. Coding all videos required 66 hours.

\paragraph{Exploratory analysis}

We perform the exploratory data analysis (EDA) using R version 3.3.2 \cite{r-project} and its R Markdown language to document every step and result of this phase. We employ the most common tools of EDA: box plots, bar charts, heat maps and summarizing tables with aggregated data.

\paragraph{Quantitative analysis}

During the data analysis, we did not make a priori assumptions about the distribution of the results. Furthermore, we performed checks for normality of the distributions that yielded negative results, thus we used non-parametric tests~\cite{hitchhiker}. In case of two of sample groups, we employed the Mann-Whitney U test. This test checks whether two groups of independent samples are from the same population ($H_0$) or not ($H_1$). It is a widely-used practice to calculate effect sizes for statistical samples and tests \cite{effect-size-estimates}. We chose one of the most prevalent metric, the Vargha-Delaney $\hat{A}_{12}$ measure to calculate and report effect sizes of Mann-Whitney U test.

For analyses, where we had to deal with more than two independent sample groups, we used the Kruskal-Wallis H test, which is the extension of the Mann-Whitney U test for exactly these cases. Its null hypothesis is that the sample groups are from identical populations, consequently the alternative hypothesis is that there is at least one sample group, which statistically dominates another. If there was a difference, we employed Mann-Whitney U test for post-hoc analysis of sample group pairs.

Note that if Mann-Whitney U Test is used for cross-checks between multiple groups, one would need to correct the significance values using e.g., Bonferroni or Dunn-Sidak correction. However, we did not used the test for these kind of checks, thus we did not perform significance correction.

\subsection{Threats to validity}

During the planning of our study, we identified the internal, external and construct threats to its validity. In terms of \emph{internal threats}, our results might be affected by the common threats of human studies~\cite{ko-guideline-experiments}. For instance, this includes the maturation effect caused by the learning of exercises, and the natural variation in human performance as well. Moreover, the students know each other and they could talk about the tasks in the study between the two sessions (see Section~\ref{sec:execution}). We eliminated this threat by using different projects and faults at each occasion. The data collection and analysis procedure might also affect the results, however we validated the video logs by R scripts and the portal functions by testing.

The \emph{generalization} of our results might be hindered by some factors. The performances of students and professional users of white-box test generators may differ. Yet, involving students is common in software engineering experiments \cite{sjoeberg-experiments-survey}, and results suggests that professional experience not necessarily increases performance \cite{professional-student-study}. Our graduate students typically have at least 6 months work experience, thus they are on the level of junior developers. Another threat to external validity is the specification given in comments, and not in a program specification. However, our goal was to carefully select open-source projects, which in general do not have formal or clear specifications of behavior. This decision on one hand may reduce the genericity of results for projects with formal specifications, but on the other hand, it increases the genericity for open-source software. Fault injection procedure could have effects on the genericity of the results, however we selected this method after thinking through several other alternatives (such as GitHub issues).

The threats to the \emph{construct validity} in our study is concerned with the independent variables. It might be the case that some of the variables we selected are not affecting the difficulty of classification of generated white-box tests. We addressed this threat by carefully analyzing related studies and experiments in terms of design and results in order to obtain the most representative set of variables.


\section{Execution}
\label{sec:execution}


\paragraph{Pilots} Our study was evaluated by two separate pilot sessions. First, we performed the tasks using ourselves as participants. After fixing the discovered issues of the design, we chose 4 PhD students having similar knowledge and experience as our intended participants to conduct a pilot in the live environment. We refined the study design based on the feedback collected (see object selection and project selection in Section~\ref{sec:experiment-design}).

\paragraph{Sessions} We separated our live study into two different sessions. On the first occasion the NBitcoin project, on the second one Math.NET was used. The sessions were carried out on 1st and 8th December 2016. Both sessions followed the procedure and could fit in the 2-hour slot.

\paragraph{Participants}
Altogether 54 students volunteered of the 120 attending the course: 30 came to the first occasion (NBitcoin) and 24 for the second (Math.NET). 34 of the students had 4 years or more programming experience, while 31 participants had at least 6 months industrial work experience. They scored 4.4 out of 5 points on average on the testing quiz of the background questionnaire.

\paragraph{Data collection and validation} 
We noticed three issues during the live sessions. In the first session, Visual Studio cached the last opened window, thus participants got three windows opened on different tabs when they started Visual Studio. In the second session, we omitted the addition of a file to the test project of Math.NET that led to 3 missing generated test cases in Visual Studio (for method \verb|CombinationsWithRepetition|). We overcame this issue by guiding the participants step-by-step on how to add that test file. This guided part lasted around 9 minutes, thus we extended the deadline to 69 minutes in that session. Finally, unexpected shutdown of two computers caused missing timing data for the first two tests for two participants (ID: 55 and 59). The rest of their experiments were recorded successfully. The experiment portal has a continuous saving mechanism, therefore their classification answers were preserved. We took all these issues into account in the timing analysis. During the data validation of the recorded data we discovered only one issue. The network in the lab room went off on 1st December, and due to this the experiment portal was not able to detect every activity. This data was recovered with the coding of the videos for each participant.


\section{Results}
\label{sec:results}


\subsection{RQ1: Performance in classification}

\begin{figure*}[ht]
	\centering
	\subfloat[NBitcoin]{\includegraphics[width=1.05\columnwidth]{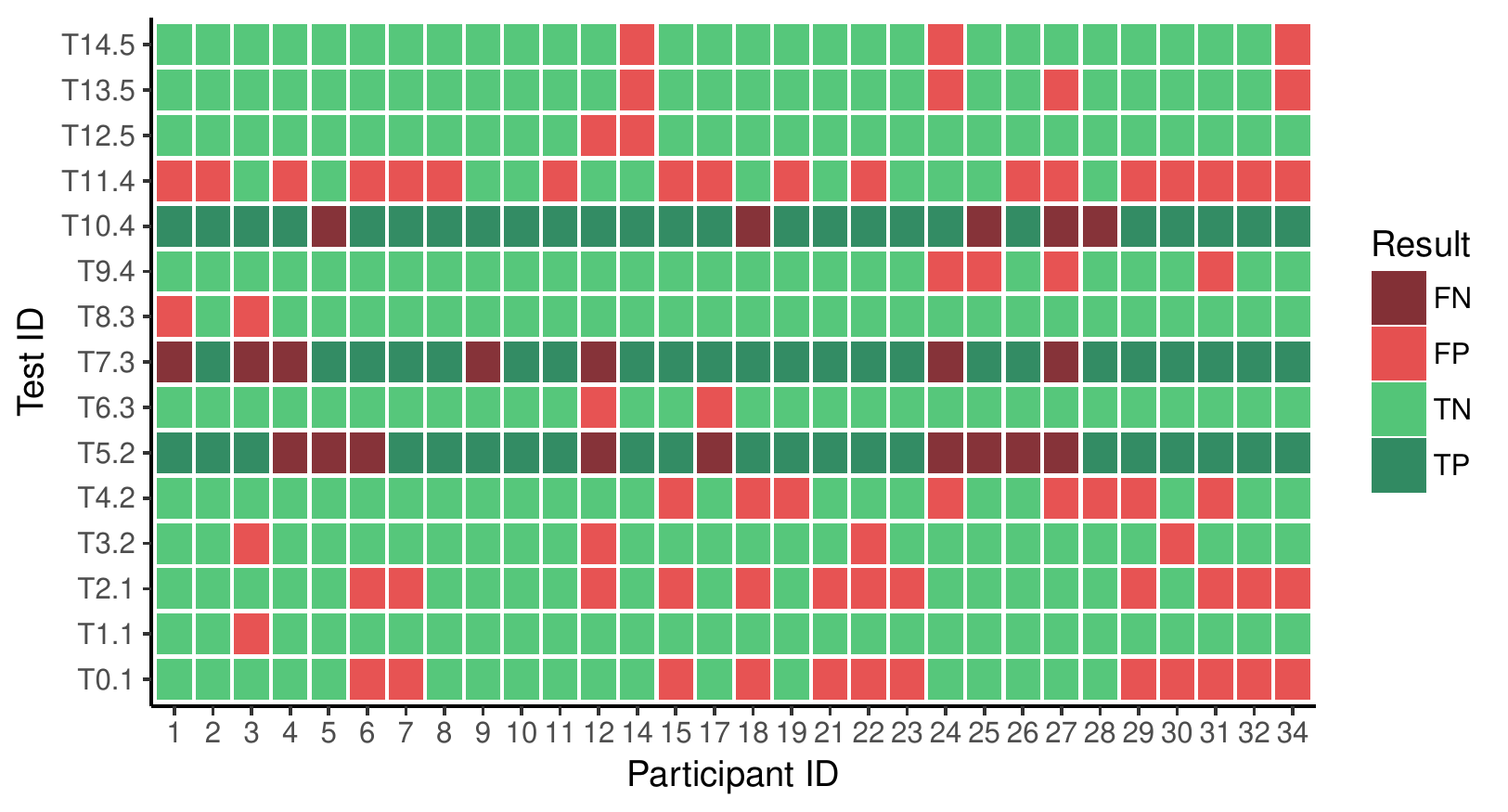}} 
	\subfloat[Math.NET]{\includegraphics[width=0.9\columnwidth]{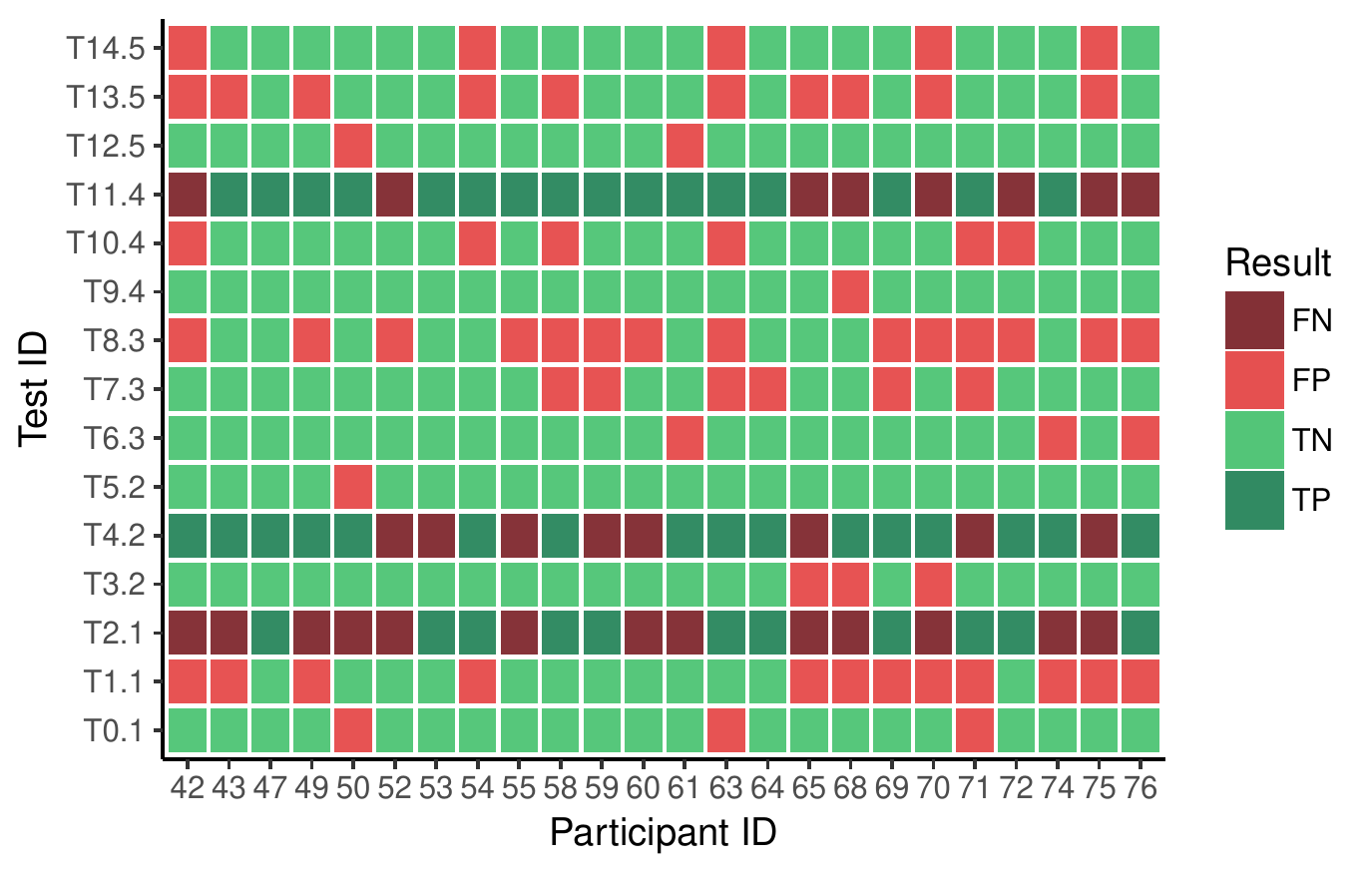}}
	\caption{Overall results of the participants measured with the common binary classification measures.}
	\label{fig:overall-results}
\end{figure*}

To evaluate the overall performance of participants in the classification of generated tests, we employed binary classification using the confusion matrix presented in Table~\ref{table:confusion-matrix}. Figure~\ref{fig:overall-results} presents the overall results. The figure encodes all four outcomes of evaluated answers. The first and foremost fact visible in the results is that there are numerous erroneous answers (marked with two shades of red). This implies that not only faulty cases were classified as fault-free, but also there were fault-free cases classified as faulty.

In case of NBitcoin, there is only one participant (ID: 10) who answered without any errors. However, there is no test, which is not marked falsely by at least one of the participants. Furthermore, one can notice two patterns in the results for NBitcoin. First, tests T0.1 and T2.1 show very similar results for the same participants. This is caused by the fact that the generated test codes and their names are very similar. However, there were no injected fault in the code, both cases encode expected behaviors with respect to the specification. The other noticeable result is that T11.4 has more false answers than true ones. This test case causes an exception to occur, yet it is an expected one. Although throwing an exception is not explicitly stated in the method comment, still the specification of the invoked and exception-causing method implies that.

In case of Math.NET, the overall results show similar characteristics: there is no test, which was correctly classified by everyone, and also there is only one participant (ID:~47) who was able to classify every test correctly. Similarly to NBitcoin, two tests show larger deviations in terms of results: T2.1 and T8.3. Taking a closer look at T2.1 (encoding a fault) reveals that its functionality was simple, participants had to examine the binomial coefficient $\binom{n}{k}$ calculation. But the fault was injected in the sanity check at the beginning of the method (this sanity check is not detailed in the specification, however, the definition of the binomial coefficient implies that). In this particular test case, the test inputs should have triggered the valid sanity check. For test T8.3, the misunderstanding could come from an implementation detail called factorial cache, which pre-calculates every factorial value from 1 to 170. The original documentation states that numbers larger than 170 will overflow, but does not detail its exact method. Test T8.3 uses 171 as input for which the implementation returns positive infinity. This is the correct behavior used consistently in the class, but probably the participants expected an overflow exception.

\begin{figure*}[ht]
	\centering
	\subfloat[True positive rate]{\includegraphics[width=0.3\textwidth]{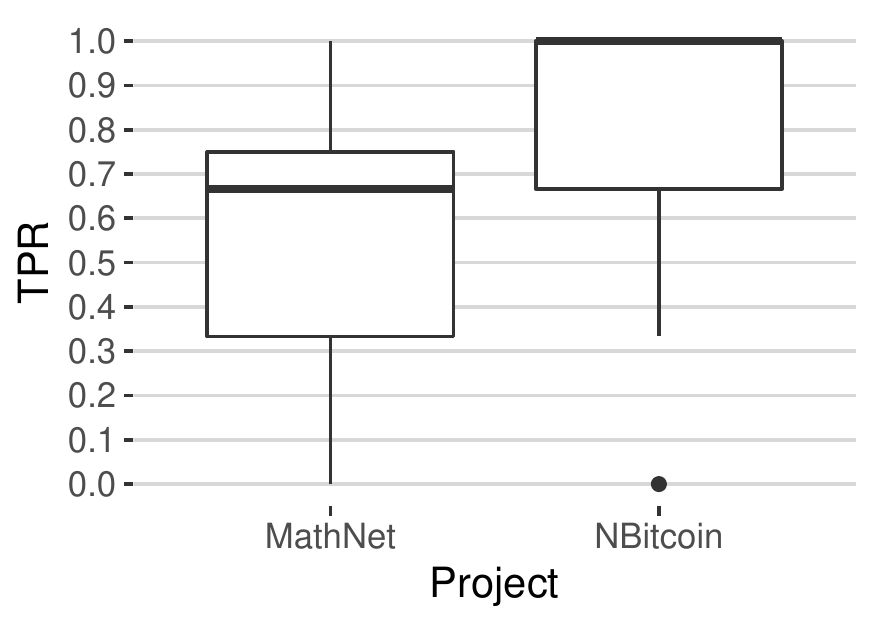}} 
	\subfloat[True negative rate]{\includegraphics[width=0.3\textwidth]{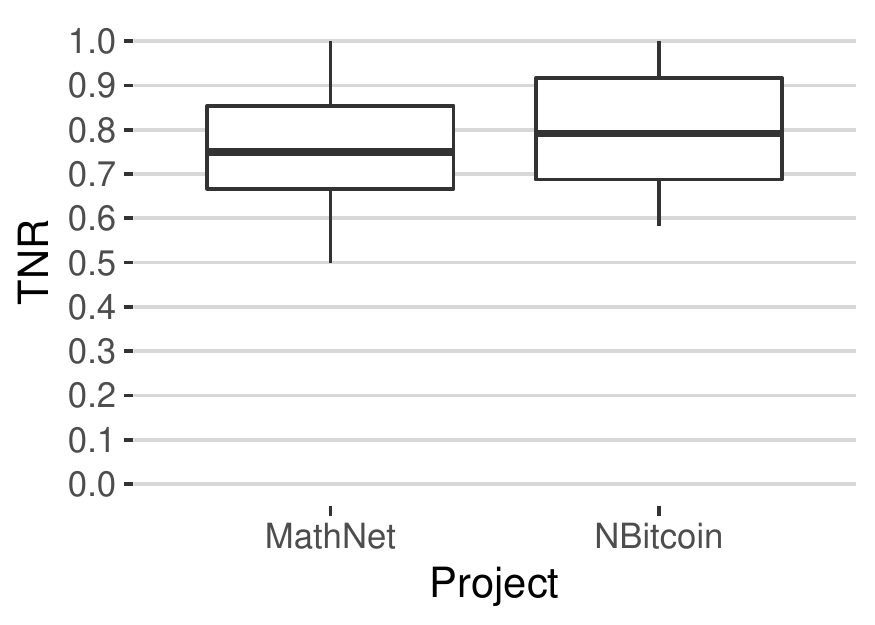}}
	\subfloat[Matthews correlation coefficient]{\includegraphics[width=0.3\textwidth]{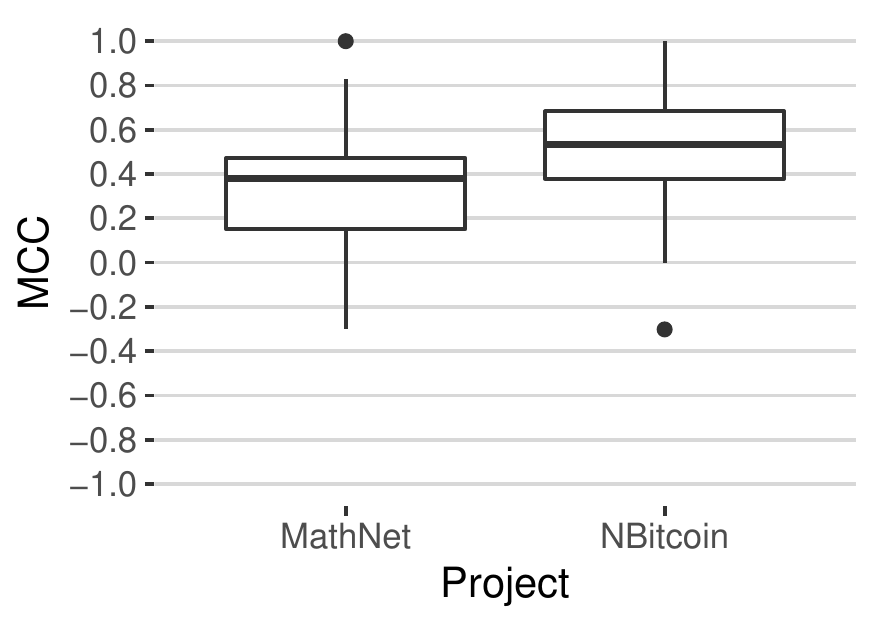}}
	\caption{Box plots of the results containing detailed metrics of binary classification.}
	\label{fig:results-boxplots}
\end{figure*}

We also analyzed the data in terms of different metrics for binary classification. The most widely used ones that are independent from the number of positive and negative samples are: true positive rate (TPR), true negative rate (TNR) and Matthews correlation coefficient (MCC) \cite{powers-matthews}. Summary of these metrics are shown in Figure~\ref{fig:results-boxplots}.

In terms of TPR, participants of the NBitcoin session outperformed the results of participants working with Math.NET. For NBitcoin, the median is 1, which means that more than half of the participants were able classify all fault-encoding tests as faulty. In contrast, results for Math.NET show that the upper quartile starts from 0.75, which is much lower.

For TNR, the two projects show very similar results with almost the same medians and inter-quartile ranges. Only a slightly wider distribution is visible for NBitcoin. This and the results for TPR confirms that the classification was easier for NBitcoin.

MCC is basically a correlation metric between the given and the true answers, and thus gives a value between -1 and 1. If MCC is zero, then the given classification has no relationship with the true classification. For NBitcoin, the MCC values show worse results than what can be expected from TPR and TNR values. The median is only around 0.55, which is only a moderate correlation. In case of Math.NET, the inter-quartile range is between 0.5 and 0.2, which can be considered as a low correlation between the true classification and the ones given by participants. Another interesting note that both experiment sessions had participants with negative correlation, which indicates that the given participant had more false answers than true ones.

\paragraph{Summary} The overall results of the participants showed a moderate classification capability. Many of them committed errors among their answers. Some of these errors were possibly caused by misunderstanding the specification, however, a large portion of wrong answers may have been caused by the difficulty of the problem.

\subsection{RQ2: Time spent for classification}

\begin{figure*}[ht]
    \centering
    \subfloat[NBitcoin]{\includegraphics[width=0.8\columnwidth]{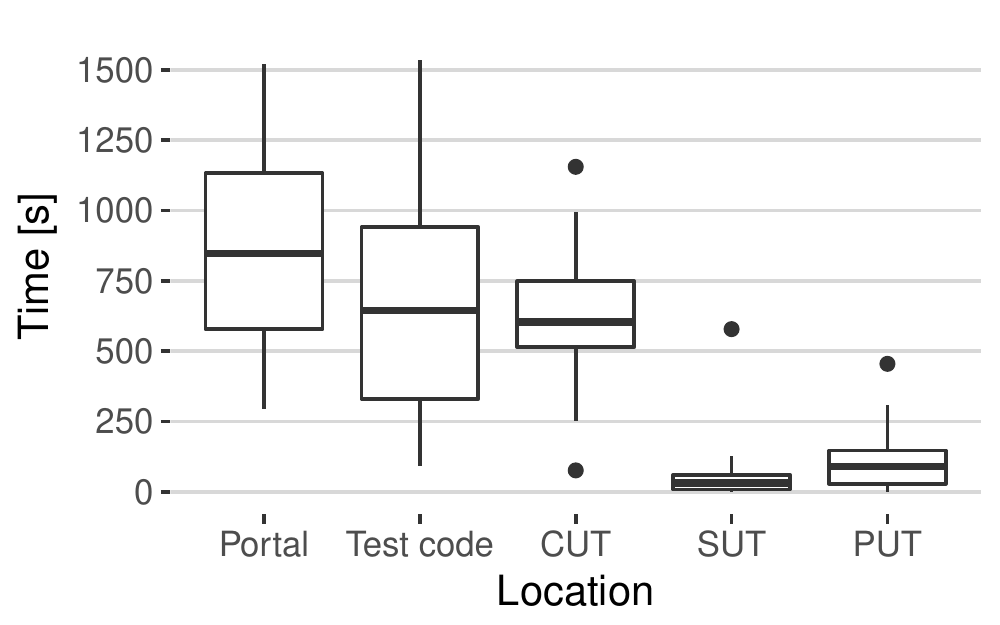}}
    \subfloat[Math.NET]{\includegraphics[width=0.8\columnwidth]{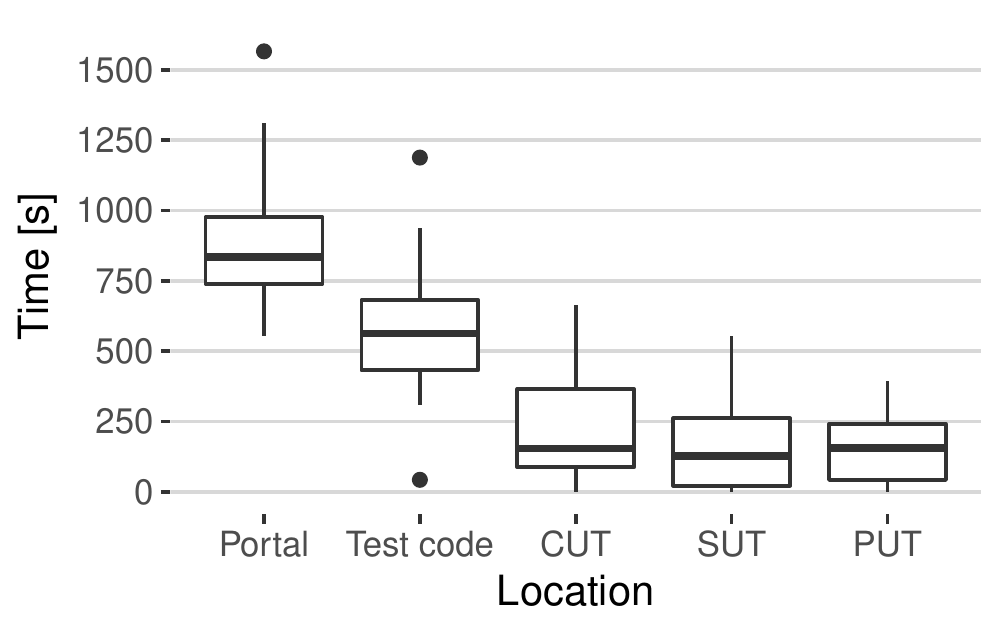}}
    \caption{Time spent on each of the possible locations.}
    \label{fig:time-spent-full}
\end{figure*}

\begin{figure*}[ht]
    \centering
    \subfloat[NBitcoin]{\includegraphics[width=\columnwidth]{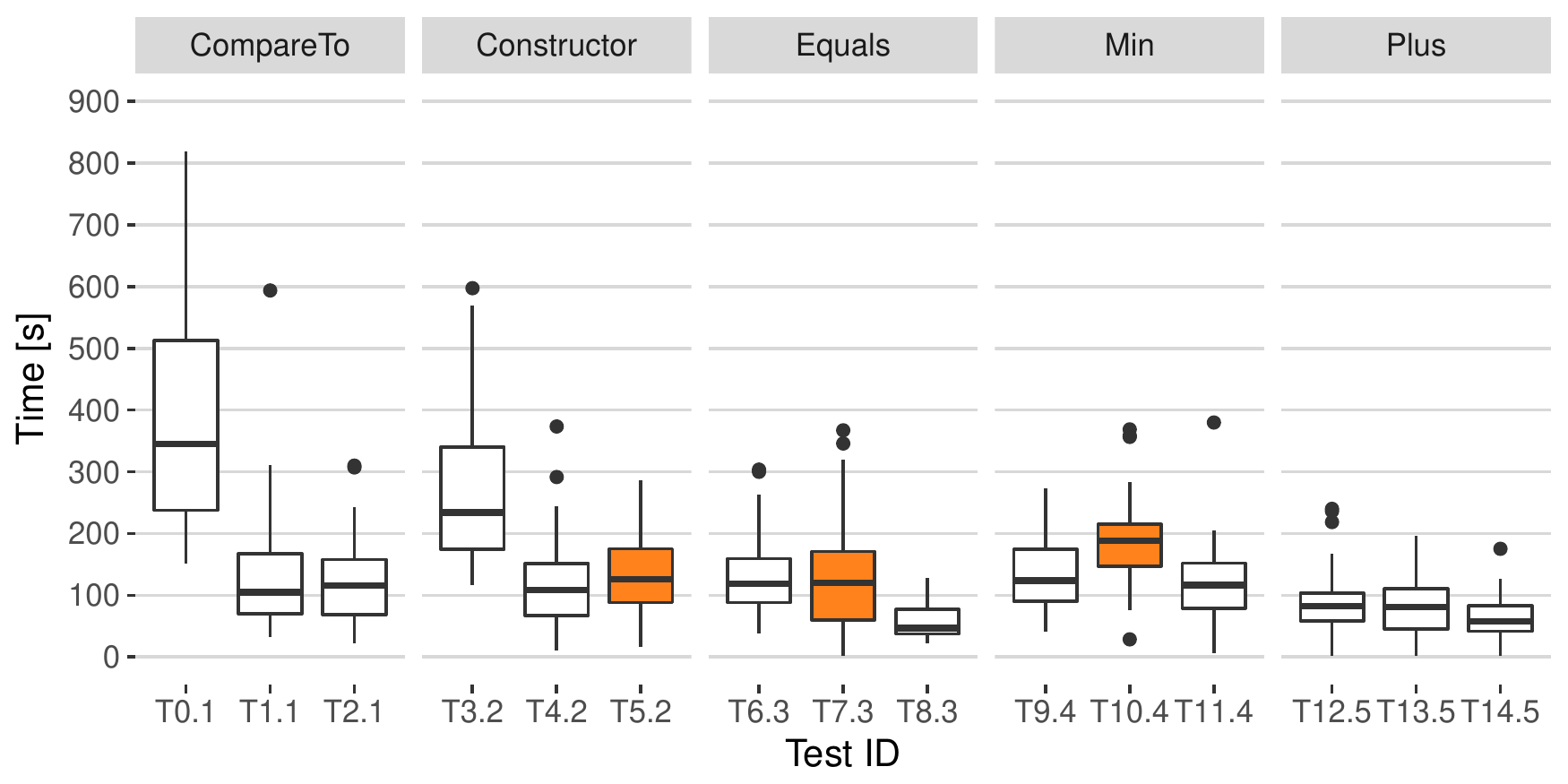}} 
    \subfloat[Math.NET]{\includegraphics[width=\columnwidth]{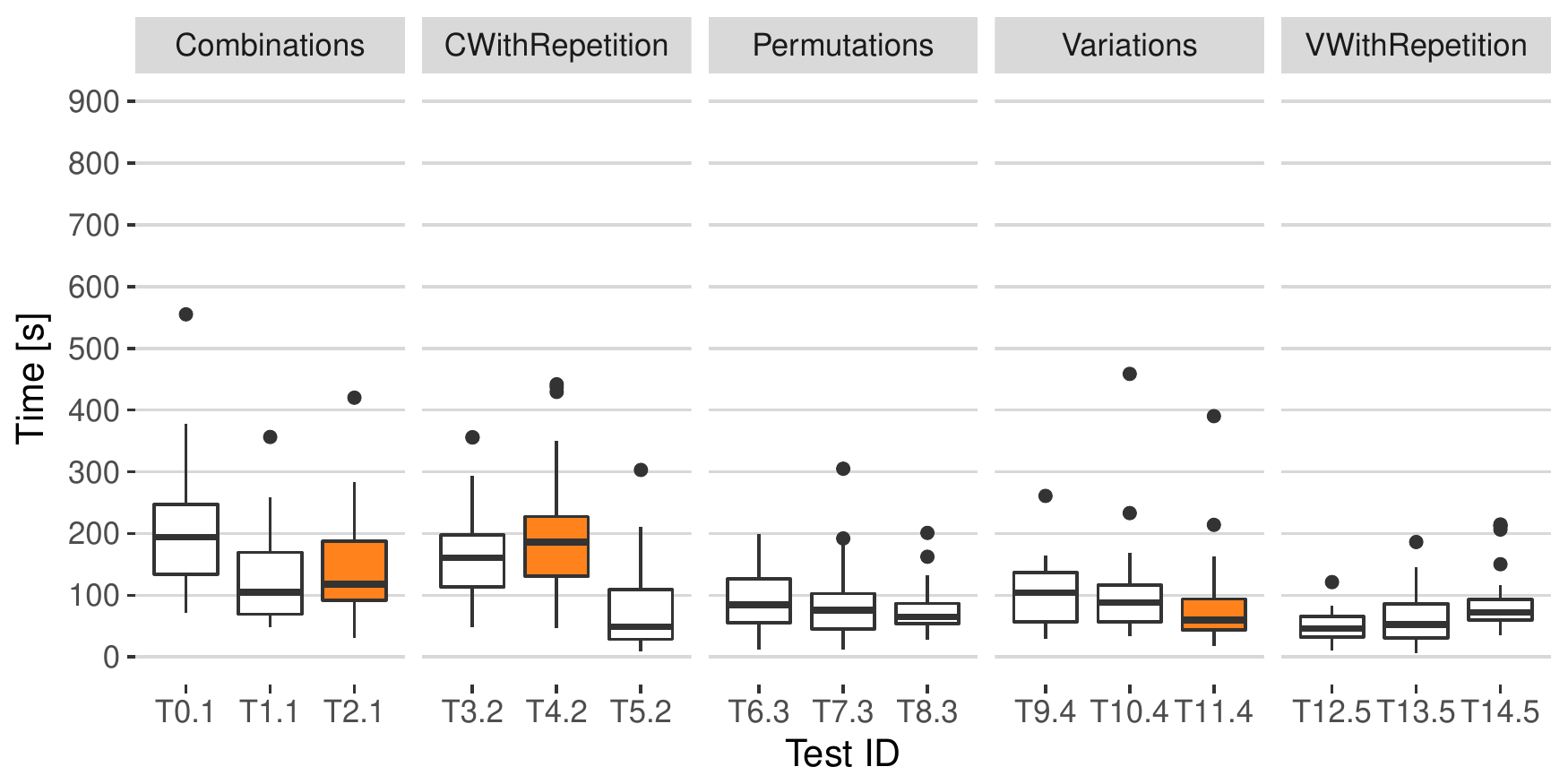}}
    \caption{Time spent in each test case per project (faulty cases marked with orange).}
    \label{fig:time-spent-tests}
\end{figure*}

\begin{table}[ht]
    \centering
    \setlength{\tabcolsep}{3pt}
    \caption{Descriptive statistics of time spent by the participants during the whole session [mins] and on each test [s].}
    \label{table:overview-descriptive}    
    \begin{tabular*}{\columnwidth}{l | G F F F F F}\toprule 
        & \emph{Project} & \emph{Min} & \emph{Median} & \emph{Mean} & \emph{Max} & \emph{sd} \\
        \midrule
        \multirow{2}{*}{\rotatebox[origin=c]{90}{\parbox[c]{0.7cm}{\small\centering Total\\$[$min$]$ }}} & NBitcoin & 37.41 & 46.94 & 46.26 & 54.90 & 3.96 \\
        & Math.NET & 34.88 & 44.57 & 44.52 & 52.61 & 5.24 \\ 
        \midrule
        \multirow{2}{*}{\rotatebox[origin=c]{90}{\parbox[c]{0.7cm}{\small\centering Tests\\$[$s$]$}}} & NBitcoin & 1.82 & 117.82 & 146.87 &  818.12 & 121.89 \\
        & Math.NET & 6.07 & 86.24 & 113.90  & 555.03 & 87.50 \\
        \bottomrule
    \end{tabular*}
\end{table}

We analyzed the data obtained from the video annotations from different aspects to have an overview of the time management of participants. Note that during the time analysis we excluded the data points of participants 55 and 59, who had missing time values for T0.1 and T1.1, as these may affect the outcome of the results.

Table~\ref{table:overview-descriptive} summarizes the total time and time spent on one test. Total time was calculated using the length of the recorded videos. For the test cases, we summed the time spent in the IDE on a specific test case and the time spent on the portal page of the given test. The total time spent during the sessions is very similar for the two project. There is a roughly 17 minutes difference between the fastest and slowest participants, while the average participant required 45 and 46 minutes to finish the classification. Note that this involves every activity including the understanding of the code under test. The results show rather large deviations in the values for the test cases. The minimum in case of NBitcoin was probably caused by two factors. First, there were participants who gained understanding of the code under test, thus were able to quickly decide on some of the tests. Second, each method had 3 test cases, and the third cases could be classified in a shorter amount of time, which emphasizes a presence of a learning curve. In contrast, participants required a rather long time period to classify some of the test cases. A rough estimation for the time required for classification based on our results is around 100 seconds.

To understand how participants managed their time budget, we analyzed their time spent on each of the possible locations (Figure~\ref{fig:time-spent-full}). These locations are the followings: portal pages of the tests, the Visual Studio windows including the test codes, class under test (CUT), other system under test (SUT) than CUT, and parameterized unit test (PUT). Note that we excluded the home page of the portal from this analysis, as it contains only a list of the cases, thus served only for navigation. The results are similar for both projects, yet there are two differences to mention. It is clear that participants mostly used the test code and the corresponding specification in the portal and in Visual Studio to understand the behavior. However, in case of NBitcoin they analyzed the class under test almost as much as the test code in Visual Studio. This is not the case for Math.NET, probably because  participants were already familiar with the domain knowledge of the tested class. Another difference is for the time spent with the system under test. Math.NET had its faults injected outside the class under test, and participants had to explore its dependencies to understand what causes the mismatch of behavior with respect to the specification.

In order to gain deeper insights into the time budget, we analyzed the time required for each test case (Figure~\ref{fig:time-spent-tests}). We calculated this metric by summarizing 5 related values: the time spent in the portal page of the test, the time spent in the Visual Studio window of the test, the time spent with CUT (class under test), PUT (parameterized unit test) and SUT (other system under test than CUT) for the test case currently opened in the portal. On a high-level overview, two trends can be noticed in the values. The first one is the decreasing amount of time required as participants progressed. The second factor is the first-test effect causing the first test to have higher values for several methods.

\paragraph{Summary} The analysis of the time spent by participants pointed out that they spend roughly around 100 seconds on average only with the test code to classify a particular generated white-box test. The time spent with other parts of the code is added on top of this. Based on the results, the users of white-box test generators may have to spend a noticeable amount of time to classify the generated tests based on their correctness.

\subsection{RQ3: Impacting factors of classification}

In order to have a better understanding on what could impact the classification performance of participants, we applied statistical methods for different aspects of the dataset. Our goal is 1) to provide information about the potential relationships between various attributes, 2) to gain important insights to the data  and 3) to define recommendations for future studies. We selected the true positive rate (TP -- sensitivity) and true negative rate (TN -- specificity) metrics to evaluate the performance of classification.

\begin{table}[ht]
	\centering
    \setlength{\tabcolsep}{3pt}
   	\caption{Impact of project selection on TP and TN rate.}
    \label{table:projects-classification}    
	\begin{tabular*}{\columnwidth}{ F G G F F F}\toprule 
	& \emph{NBitcoin mean} & \emph{MathNet mean} & \emph{sd} & \emph{U pv} & \emph{$\hat{A}_{12}$} \\
	\midrule
	TP rate & 0.767 & 0.597 & 0.309 & 0.045 & 0.651 \\
	TN rate & 0.800 & 0.771 & 0.130 & 0.473 & 0.557 \\
	\bottomrule
	\end{tabular*}
\end{table}

\paragraph{Project selection} Foremost, we analyzed whether the project selection has influence on the classification performance. As there were two groups of samples (NBitcoin with 30 and MathNet with 24 samples), we used the Mann-Whitney U test. The significance of the test (U pv) along with the means, effect sizes ($\hat{A}_{12}$) and the mean of standard deviations (sd) are shown in Table~\ref{table:projects-classification}. In case of TP rate, the values have reasonably large differences among the two projects (this can be also seen on Figure~\ref{fig:results-boxplots}). Based on the p-value of the Mann-Whitney U test, one can reject the null hypothesis that the values are from the same population with 95\% confidence. Also, the Vargha-Delaney metric has a value of 0.651, which is considered as a medium difference between the two sample groups. In contrast, the p-value for TN rate shows there is not enough statistical significance to state that the two groups of samples are from different populations. The $\hat{A}_{12}$ value also supports that they are similar, as 0.557 is considered as a small difference between the values in the sample groups.

\paragraph{Programming experience} Participants filled the background questionnaire prior to the experiment. We had several questions on their experiences, one of them was regarding their programming experience measured in years. We defined 7 levels in the survey from which they only selected 6. However, only 1 participant selected having no experience (possibly omitted answering) and 2 participants selected less than 1 year. We excluded their answers as they form a very small group to be used in a statistical test. Thus, in the end, we had 51 samples with 4 levels of programming experience: 2 years (N=9), 3 years (N=8), 4 years (N=19), 5 or more years (N=15). In order to determine if there is any difference between the sample groups, we used the Kruskal-Wallis H test. The results are shown in Table~\ref{table:progexp-classification}. In case of true positive rate, the p-value of the Kruskal-Wallis is 0.122, which means that one cannot significantly reject the null hypothesis, thus the groups are from rather likely to be from identical populations. Similarly for true negative rate, the p-value is 0.182, which yields the same results as for TP rate.

\begin{table}[ht]
    \centering
    \setlength{\tabcolsep}{3pt}
    \caption{Impact of programming experience on TP and TN rate.}
    \label{table:progexp-classification}    
    \begin{tabular*}{\columnwidth}{ F F F F F F F}\toprule 
        & \multicolumn{4}{c}{\emph{Means per years}} & & \\
        \cmidrule{2-5}
        & \emph{2} & \emph{3} & \emph{4} & \emph{5+} & \emph{sd} & \emph{H pv} \\
        \midrule
        TP rate & 0.592 & 0.542 & 0.807 & 0.667 & 0.316 & 0.122 \\
        TN rate & 0.703 & 0.802 & 0.816 & 0.811 & 0.123 & 0.182 \\
        \bottomrule
    \end{tabular*}
\end{table}

\begin{table}[ht]
	\centering
    \setlength{\tabcolsep}{3pt}
   	\caption{Impact of industrial experience on TP and TN rate.}
    \label{table:indexp-classification}    
	\begin{tabular*}{\columnwidth}{l H H H H H H H}\toprule 
	& \multicolumn{5}{c}{\emph{Means per experience}} & & \\
	\cmidrule{2-6}
	& \emph{None} & \emph{$<$6 months} & \emph{7-12 months} & \emph{1-2 years} & \emph{3-5 years} & \emph{sd} & \emph{H pv} \\
	\midrule
	TP rate & 0.533 & 0.741 & 0.815 & 0.574 & 0.917 & 0.271 & 0.110 \\
	TN rate & 0.850 & 0.842 & 0.722 & 0.731 & 0.854 & 0.123 & 0.026 \\
	\bottomrule
	\end{tabular*}
\end{table}

\begin{table}[ht]
    \centering
    \setlength{\tabcolsep}{3pt}
    \caption{Difference in TN rate w.r.t. to work experience.}
    \label{table:indexp-class-1-3}    
    \begin{tabular*}{\columnwidth}{ F G G F F F}\toprule 
        & \emph{$<$6 months} & \emph{1-2 years} & \emph{sd} & \emph{U pv} & \emph{$\hat{A}_{12}$} \\
        \midrule
        TN rate & 0.842 & 0.731 & 0.771 & 0.009 & 0.751 \\
        \bottomrule
    \end{tabular*}
\end{table}

\paragraph{Work experience} Based on the background questionnaire filled by the participants, we analyzed the connection between their classification performance and their work experience. For the question regarding the industrial work experience, we also defined 7 levels to choose from. However, the 54 participants have only selected 5 of them, which were the followings: none (N=5), less than 6 months (N=18), 7-12 months (N=9), 1-2 years (N=18) and 3-5 years (N=4). The results for this analysis are found in Table~\ref{table:indexp-classification}. Using the Kruskal-Wallis H test for the TP rate, the results show that the differences are not statistically significant enough to reject that the sample groups are from identical populations. One may note that means are very different for the groups, however the results are not significant enough to reject the null hypothesis. This may be caused both by the varying number of participants per group and by the lack of robustness in mean. An interesting phenomenon can be observed for the results in TN rate. The Kruskal-Wallis test tells that there is a statistically significant difference between some of the sample groups (rejecting $H_0$). In order to gain more information about the differences, we used the Mann-Whitney U test on the two largest sample groups (less than 6 months experience and 1-2 years of experience). The results for this test can be found in Table~\ref{table:indexp-class-1-3}. The values shows that there is a statistically significant difference between the two groups in terms of TN rate (rejecting $H_0$). Furthermore, the value of the $\hat{A}_{12}$ statistic indicated a large difference with the value of 0.751.

\paragraph{Summary} The results for RQ3 showed insights on relationships. In terms of project selection, the results may indicate that the project under test or its attributes (e.g., complexity, fault location, type of faults, etc.) has influence on true positive rate. For participants with different programming experiences, our results showed no significant differences in their classification performance. Last, we analyzed the industrial work experience of participants that showed TN rate was significantly affected by the work experience in an inverse-way for the two largest group of our participants.


\section{Discussion}
\label{sec:discussion}

\subsection{Implications of the results}

The results for RQ1 showed that classifying the correctness of generated white-box tests could be a challenging tasks. The median of misclassification rate was 33\% for fault-encoding tests, while 25\% for correct tests. Both could be caused by several factors such as 1) the misunderstanding of specification, 2) the misunderstanding of class behavior, or even 3) the underlying fault attributes in the software under test. However, most likely the challenge in classification is mostly due to the combination of these causal factors.

For RQ2, our results showed that participants could spend minutes only to understand the encoded behavior and functionality in the test cases even for the selected classes. Moreover, this does not include other time spent in the IDE, which was mostly used to understand the code under test. The results may yield that developers and testers could spend a non-negligible amount of time with the classification of test cases, which may reduce the time advantage provided by automatically generated white-box tests.

In terms of RQ3, we obtained interesting factors that may have influence on participant classification performance. First, our results showed significantly that the attributes of the code under test may influence the performance. Further studies may be required to support this hypothesis that control some attributes of code under test as independent variables. A study like this may reveal, which attribute has the most influence in performance. Furthermore, our results had significance in the analysis of relationship between the participant work experience and their classification performance. Further studies may address this topic by controlling participant industrial experience with a reasonable amount of participants.

\subsection{Insights from participants' behavior}

By watching all screen capture and performing video coding, we gained important insights about the user activities and behaviors during the classification of generated white-box tests. As expected, many participants employed debugging to examine the code under test. They mostly checked the exceptions being thrown, parameterized unit tests for the test methods, and assertions generated into the test code. This emphasizes the importance of debugging as a tool for investigating white-box test behavior. Most of the participants executed the tests to check their actual outcome. Some cases contained unexpected exceptions, which confused few participants.

Another interesting insight we obtained is that some participants spent only seconds with the examination of the last few test cases. This could point out that they either gained understanding of the code under test by the end of the session (i.e., learning factor), or they got tired by the continuous attention required during the task.

\subsection{Suggestions from exit survey}

The participants in our study filled an exit survey at the end of the sessions. They had to answer both Likert-scaled and textual questions. The results for the agreement questions yielded that participants had enough time to understand the class under test and to review the generated tests. Most of them also selected that it was easy to understand the class and the tests. They agreed that the generated tests were difficult to read, however the answers were almost equally distributed for the questions about the difficulty of the task and the confidence in their answers. This shows that they are mostly not very confident in their own answers. Furthermore, the feedback about the time and difficulty showed that our study design was appropriate in terms of these.

In their textual answers participants mentioned the difficulties in reviewing the tests and gave several suggestions to improve the test code (some of these were also reported in the literature~\cite{tillman-pex-experience}).

{\small 
    \begin{itemize}
        \item ``Deciding whether a test is OK or wrong when it tests an unspecified case. (e.g. comparing with null, or equality of null)''
        \item ``Distinguishing between the variables was difficult (assetMoney, assetMoney1, assetMoney2).''
        \item ``Tests should compare less with null and objects with themselves.''
        \item ``I think that some assertions are useless, and not asserting 'real problems', just some technical details.''
        \item ``Generated test cases doesn't seperated into Arrange, Act, Assert and should create more private methods for these concerns.''       
        \item ``Generate comments into tests describing what happening.''
    \end{itemize}
}

Our recommendation for improving test generators to help developers and testers with generated assertions consists of the followings.

\begin{itemize}
	\item Instead of using the \verb|assert| keyword, test generators shall use the \verb|observed| or \verb|likelyAssert| keywords.
	\item Generated tests having null inputs shall be distinguished from the others.
	\item Generated tests shall contain variables with more meaningful names (as already implemented in refactoring features of many IDEs).
	\item The generated tests shall employ the Arrange, Act, Assert pattern in the structure of generated tests.
	\item The tests shall contain intra-line comments that describe what the given line is responsible for.
\end{itemize}


\section{Conclusions}
\label{sec:conclusions}

This paper presented an exploratory study on whether developers could validate generated white-box tests. The study performed in a laboratory setting with 54 graduate students resembled a scenario where junior developers having a basic understanding of test generation had to test a class in a larger, unknown project with the help of a test generator tool. The data showed that participants incorrectly classified a large number of both fault-encoding and correct tests (with median misclassification rate 33\% and 25\% respectively). The results confirm the findings of previous studies and broaden their validity. The implication of the results is that the actual fault-finding capabilities of the test generator tools could be much lower than reported in technology-focused experiments. Thus we suggest to take into account this factor in future studies.

An experimental study always has limitations. We collected important context variables that could affect the classification performance (e.g., experience, source code access), and defined the levels chosen in the current study that collectively reflect one possible scenario. As in our study all variables had fixed levels, this naturally limits its validity. Future studies altering these settings could help to build a ``body of knowledge''~\cite{basili-families-of-epxeriments}. Our analysis indicates that the object under study and the participants' industrial experience could be possible factors. Moreover, designing a study where participants work on a known project or perform regression testing would be important future work. Therefore we made available our full dataset, coded videos and lab package to support further analyses or replications.

\bibliographystyle{ACM-Reference-Format}

\bibliography{bib/references}

\end{document}